\documentclass[aps,prl,reprint,twocolumn,amsmath,amssymb,citeautoscript,longbibliography]{revtex4-1}
\usepackage{graphicx}
\usepackage{appendix}
\usepackage{csquotes}
\usepackage{dcolumn}
\usepackage{bm}
\usepackage{latexsym,epsfig}
\usepackage{graphicx}
\usepackage{verbatim}
\usepackage{comment}
\usepackage{amsmath}
\usepackage{amssymb}
\usepackage{physics}
\usepackage{stmaryrd}
\usepackage{color}
\usepackage{epstopdf}
\usepackage{grffile}
\usepackage{lipsum}
\usepackage{enumitem}
\usepackage{ulem}

\usepackage[dvipsnames]{xcolor}
\DeclareGraphicsExtensions{.eps}

\newcommand{\beq}{\begin{equation}}
\newcommand{\eeq}{\end{equation}}
\newcommand{\bea}{\begin{eqnarray}}
\newcommand{\eea}{\end{eqnarray}}
\newcommand{\ben}{\begin{eqnarray*}}
\newcommand{\een}{\end{eqnarray*}}
\newcommand{\bfig}{\begin{figure}}
\newcommand{\efig}{\end{figure}}

\usepackage{hyperref}
\hypersetup{
    colorlinks=true,      
    urlcolor=blue,
    citecolor=blue,
    linkcolor=blue
}

\begin{document}

\title{Emergence of distinct exact mobility edges in a quasiperiodic chain}

\author{Sanchayan Banerjee$^{1,2}$, Soumya Ranjan Padhi$^{1,2}$ and Tapan Mishra $^{1,2}$}

\email{mishratapan@niser.ac.in}

\affiliation{$^1$ School of Physical Sciences, National Institute of Science Education and Research, Jatni 752050, India}

\affiliation{$^2$ Homi Bhabha National Institute, Training School Complex, Anushaktinagar, Mumbai 400094, India}

\date{\today}

\begin{abstract}
Mobility edges (MEs) constitute the energies separating the localized states from the extended ones in disordered systems. Going beyond this conventional definition, recent proposal suggests for an ME which separates the localized and multifractal states in certain quasiperiodic systems - dubbed as the anomalous mobility edges (AMEs). In this study, we propose an exactly solvable quasiperiodic system that hosts \textit{both the conventional and anomalous mobility edges}  under proper conditions. We show that with increase in quasiperiodic disorder strength, the system first undergoes a delocalization to localization transition through an ME of conventional type. Surprisingly, with further increase in disorder, we obtain that a major fraction of the localized states at the middle of the spectrum turn multifractal in nature. Such unconventional behavior in the spectrum results in two AMEs, which continue to exist even for stronger quasiperiodic disorder. We numerically obtain the signatures of the coexisting MEs complement it through analytical derivation using Avila’s global theory. In the end we provide important signatures from the wavepacket dynamics.

\end{abstract}

\maketitle

\paragraph*{Introduction.-}
\label{sec:intro}
Quasiperiodic lattices are known to exhibit delocalization-to-localization transition in the single-particle spectrum in low dimensions, distinguishing them from lattices with random disorder where an infinitesimal disorder causes localization of all the quantum states~\cite{Anderson_1958,Gang_of_four_1979}. The Aubry-Andr\'e (AA) model, a paradigmatic example describing a tight-binding lattice with an onsite quasiperiodic potential, is the simplest model to showcase localization transition at a sharp critical disorder strength~\cite{Aubry_1980}. 
However, deviations from the tight-binding limit or generalizations of the quasiperiodic potential result in a delocalization-localization transition through some intermediate phases hosting both the localized and delocalized states separated by energy-dependent mobility edges (MEs)~\cite{ME_bich_sarma_2017, ME_Incomm_Opt_sarma2010, Roy_prl_2021, Pedro_2023, PRB_GME_sarma_2023,Loc_AA_non-nerest, Santos_prl_2019, Auditya_2021,Tanay_PRR_2020, Tanay_PRE_2020, Sarma1988, Sarma2009, Holthaus2007, Wang2020PRL,Archak,Chiaracane_2020,Yamamoto_2017,Whitney_2014}. 
Advances in experimental platforms have enabled the observation of these phenomena across diverse systems, including photonic lattices~\cite{GAO202558, Lahini_2009_prl, Elshaari_mosaic, Rechtsman_2023_reentarnt}, electric circuits~\cite{Ganguly2023, Zhang_PrB_circuit}, ultracold atoms~\cite{expt_Int_ME_GAA, Ozawa2019, Meier2016, Gadway2015, Immanuel_prl_2018}, and Rydberg atom arrays~\cite{yoshida2024proposalexperimentalrealizationquantum} and superconducting circuits~\cite{Roushan2017, Li2023}.

Adding to this rich spectral landscape, the existence of multifractal states and critical phases has been established in quasiperiodic systems~\cite{nilanjan, 2dcritical, SantosPhysRevLett, MoessnerSciPostPhys, Dancan_2d, LonghiAMEs2022, PRL_alexei_AME}. 
Such critical states, which often emerge in intermediate regimes, enrich the spectral complexity of quasiperiodic systems and provide deeper insights into the interplay between localization and delocalization of the quantum states. 
One such spectral manifestation in this context is the recently proposed anomalous mobility edges (AMEs) mediating transitions between localized and multifractal states~\cite{LonghiAMEs2022, PRL_alexei_AME, Dancan_2d}. Such anomalous features in the spectrum go beyond the concept of conventional MEs and have spurred renewed interest in understanding the nature of mobility edges in quasiperiodic systems. While the conventional MEs have been predicted and observed in numerous systems, the systems possessing AMEs are still limited. Moreover, the coexistence of both ME and AME in an exactly solvable model is a highly unusual phenomenon.

In this work, we introduce an exactly solvable one dimensional quasiperiodic lattice with a generalized Aubry-André potential where we demonstrate the coexistence of both the types of MEs in the spectrum as a function of the onsite disorder strength. We show that in certain parameter regime of the generalized potential and with increase in disorder strength, an initially delocalized spectrum completely localizes through an ME as expected. Surprisingly, however, with a further increase in the disorder strength, most of the localized states from the middle of the spectrum become multifractal in nature, forming an intermediate phase that hosts two AMEs. Remarkably, these AMEs are found to be independent of the disorder strength, as a result of which the number of multifractal states lying between the two AMEs increases with increase in disorder strength. This coexistence of mobility edges also demonstrates a new type of re-entrant behavior in the spectrum where a fully localized spectrum becomes intermediate, which constitutes localized and multifractal states. We obtain these non-trivial properties by numerically solving the model and by analytically calculating the mobility edge using Avila's global theory. We also explore the signatures of these findings from the wavepacket dynamics. 

\paragraph*{Model.-}
\label{sec:model}
We consider a one-dimensional quasiperiodic model, which is slightly different from the well-known generalized Aubry-André (GAA) model proposed in ~\cite{DasSarmaPhysRevLett, DualityPhysRevB}. The Hamiltonian that describes the system under consideration is given by 
\begin{align}
	\mathcal{H} &= -J\sum_{i} \big{(} \hat{c}^\dagger_i \hat{c}_{i+1} + H.c.\big{)} \nonumber \\ &+ \sum_{i} \frac{\lambda \cos(2\pi\beta j + \phi)}{1 - \lambda\alpha \cos(2\pi\beta j + \phi)} \hat{n}_i,
 	\label{eq:Eq1}
\end{align}
where, $\hat{c}_i^\dagger (\hat{c}_i)$ denote the creation (annihilation) operator and $\hat{n}_i$ is the number operator of the particles at site $i$. The first term in the Hamiltonian represents the nearest-neighbor hopping with amplitude $J$. The second term introduces the onsite quasiperiodic disorder in the system of strength $\lambda$. Quasiperiodicity is ensured by defining $\beta = (\sqrt{5} - 1)/2$, the inverse golden ratio, which is an irrational number~\cite{aubry1980analyticity, Harper1955, Thouless1983, B2, B3, B4, B5, B6, B7, B8, B9, B10, B11, B12, beta2, beta3}. $\phi$ is the phase of the onsite potential. The parameter $\lambda\alpha$ defines the generalization of the quasiperiodic potential. Note that compared to the original GAA model, here the generalization parameter depends on the strength of the disorder potential $\lambda$. For our analysis, we fix $J = 1$ as the energy unit and assume periodic boundary conditions (PBC) unless otherwise specified. The numerical calculations are performed by exact diagonalization of the model shown in Eq.~\ref{eq:Eq1}.

\paragraph*{Results.-}
\label{sec:Results}
We first discuss our central results i.e., the coexisting mobility edges in the spectrum.
As already mentioned before, when $\alpha = 0$, the model reduces to the AA model which exhibits a sharp delocalization-localization transition (no ME) across the entire spectrum at $\lambda = 2J$ due to the self-duality of the model. However, we show that with increase in the value of $\alpha$, the system exhibits much richer phenomena as a function of $\lambda$. As a first step in this direction, we explore the nature of the states in the spectrum. Primarily, the localization and delocalization properties of the states can be understood from the inverse participation ratio (IPR) and the normalized participation ratio (NPR) for the $m^{th}$ eigenstate, which are defined as follows, 
$
\text{IPR}^{(m)} = \sum_{i=1}^{L} |\psi_i^{(m)}|^4
, \quad \text{NPR}^{(m)} = L \left(\sum_{i=1}^{L} |\psi_i^{(m)}|^4\right)^{-1}$,  
where $i$ denotes the site index. 
Specifically, $\text{IPR}^{(m)} = 0$ (with $\text{NPR}^{(m)}$ finite) indicates a delocalized state, while $\text{IPR}^{(m)} \neq 0$ (with $\text{NPR}^{(m)} = 0$) corresponds to a localized state in the thermodynamic limit~\cite{Paredesreview_2019}. While the IPR and the NPR together can identify the localized or delocalized nature of the states, they fail to capture the information about the states which are multifractal. In this circumstance, the fractal dimension serves as an important diagnostics for a clear distinction of the different states. The fractal dimension is extracted from the generalized inverse participation ratio (IPR)~\cite{Fraxnet2022, Roy2022, Sthitadhi2018, Evers2008} and its scaling exponent $\tau_q$, defined by the relation:
\begin{equation}
    \text{IPR}_q^{(m)} = \sum_{i=1}^{L} |\psi_i^{(m)}|^{2q} \propto L^{-\tau_q},
\end{equation}
where $\tau_q$ is the mass exponent and $q$ is a real number. For localized states, $\tau_q$ vanishes, while for delocalized states, it varies linearly with the system dimension $d$ as $\tau_q = d(q - 1)$. For multifractal states, however, $\tau_q$ exhibits a nonlinear dependence on $q$, which can be expressed as: $\tau_q = D_q (q - 1)$, where $D_q$ represents the fractal dimension of the eigenstates. Specifically, a delocalized state has $D_q = 1$, a localized state has $D_q = 0$, and intermediate values of $D_q$ (between 0 and 1) signify the fractal nature of the eigenstates. In our analysis, we focus on the fractal dimension $D_2$, corresponding to $q = 2$, which is obtained from the scaling relation
$\text{IPR}_2 \propto L^{-D_2}$.

\begin{figure}[!t]
\centering
\includegraphics[width=1.0\columnwidth]{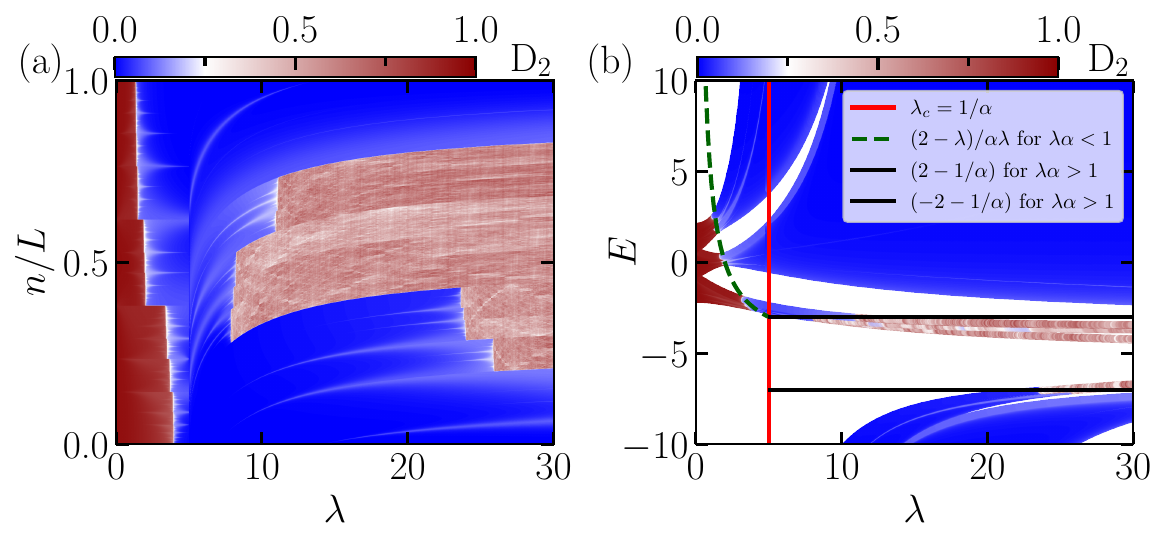}
\caption{(a) Eigenstate index fraction ($n/L$) vs. $\lambda$ for $\alpha = 0.2$ and  (b) Energy eigenvalues Vs. $\lambda$ with $D_2$n values of all the states (color-coded) indicating the nature of the states. The analytically obtained curves in (b) represent the ME (green dashed line) and the AMEs (black solid lines). All calculations are performed for a system size $L = 6765$ under periodic boundary conditions (PBC).}
\label{fig:fig1}
\end{figure}

In Fig.~\ref{fig:fig1}(a) we plot the values of $D_2$ as a function of eigenstates index fraction $n/L$ and $\lambda$. We obtain that the spectrum is entirely delocalized for $0<\lambda<1.325$ ($D_2=1$, red region). With increase in $\lambda$, the delocalized states gradually start to localize, and eventually all the entire spectrum becomes localized ($D_2=0$, blue region). This delocalization to localization transition happens through an intermediate phase hosting an ME  of conventional type (line separating the red and blue region). Interestingly, with a further increase in $\lambda$, a large fraction of the states in the spectrum become multifractal in nature, which are identified by their corresponding $D_2$ values, which lie within $0<D_2<1$ (light red region). Such multifractal state lie at the middle of the spectrum sandwiched between the localized state seen as the blue regions at the top and the bottom of the spectrum. As a result of this we obtain two AMEs at the interface of localized and multifractal states. Moreover, we obtain that the number of states becoming multifractal also increases with increase in $\lambda$.

\begin{figure}[!t]
\centering
\includegraphics[width=1.0\columnwidth]{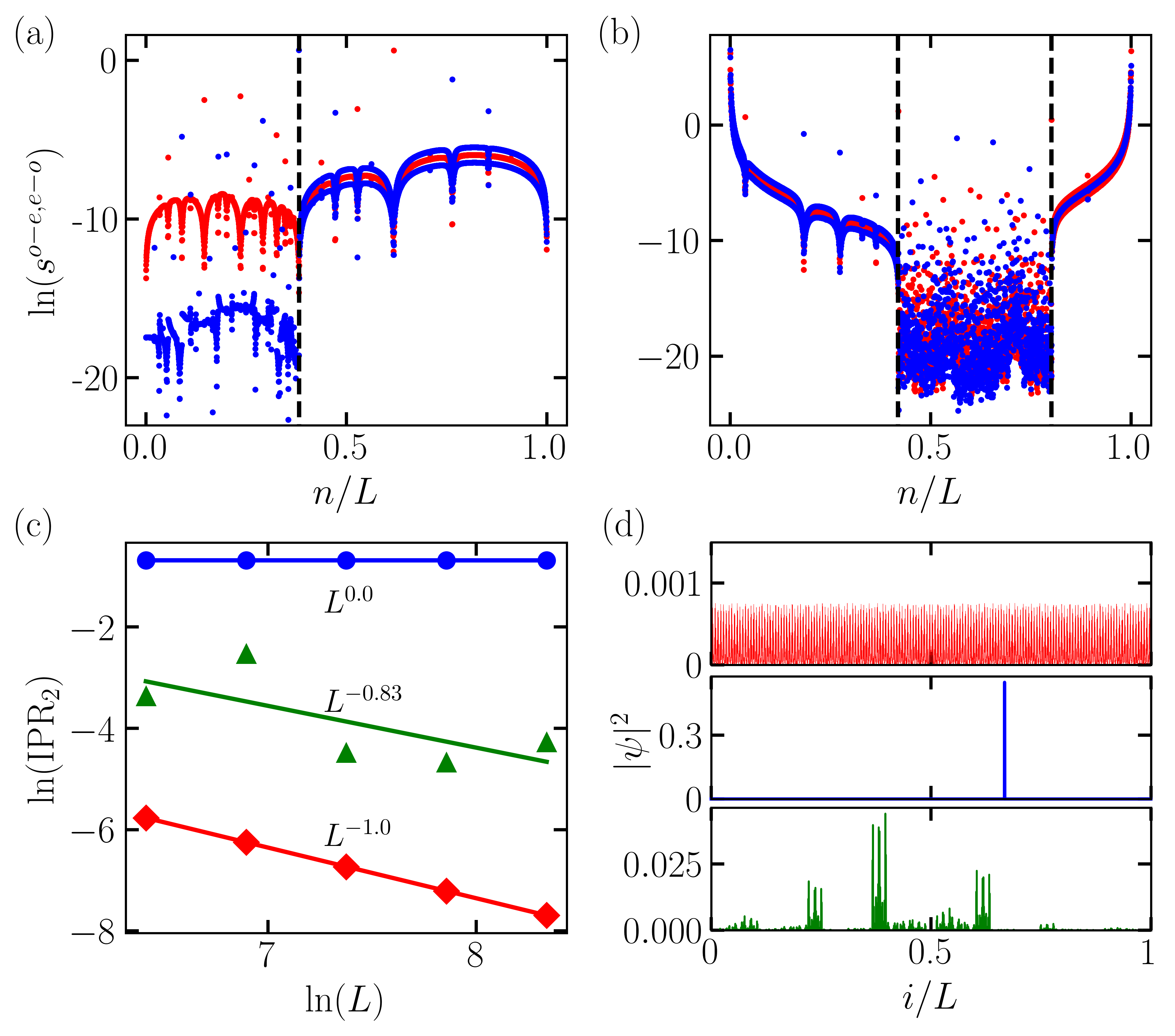}
\caption{(a, b) Level spacing for odd-even $(s^{o-e})$ and even-odd  $(s^{e-o})$ as functions of the eigenstate index fraction $(n/L)$ for $\lambda = 3$ and $\lambda = 20$, respectively (see text for details) for system size $L=6765$. (c) $\ln(\text{IPR}_2)$ is plotted as a function of $\ln(L)$, where the slopes of the straight line correspond to mass exponents of $0$, $-0.83$, and $-1$ for delocalized (red diamonds), multifractal (green triangles), and localized states (blue circles), respectively for the system size $L=610, 987, 1597, 2584, 4181$. (d) Probability density as a function of site index fraction $(i/L)$, illustrating the spatial characteristics of delocalized (top), localized (middle), and multifractal states (bottom) for $\lambda = 1,3,20$ respectively. Here, $L=4181$ is shown for clarity, and the corresponding state index is 2424.}
\label{fig:fig2}
\end{figure}

To further quantify the nature of these states, we analyze the level spacing distribution using eigenenergies $E_i$. Specifically, we define even-odd spacings $s_i^{e-o} = E_{2i} - E_{2i-1}$ and odd-even spacings $s_i^{o-e} = E_{2i+1} - E_{2i}$. This definition ensures that while for the delocalized states a doubly degenerate spectrum ($s_i^{o-e} \approx 0$), generates a gap between $s_i^{e-o}$ and $s_i^{o-e}$,  for localized states, both subsets of spacings are similar, and the gap vanishes~\cite{Santos_prl_2019, Fraxnet2022}. This behavior is illustrated in Fig.~\ref{fig:fig2}(a), where we plot $\ln(s_{e-o}, s_{o-e})$ versus the fraction of eigenstate indices $n/L$ for $\lambda=3$, clearly showing the mobility edge separating delocalized and localized states (marked by a vertical dashed line). Next, we turn our attention to the intermediate region at $\lambda = 20$, where two AMEs (marked by two dashed vertical lines) separate multifractal states at the middle of the spectrum from the localized states. The level spacing analysis, in this case, reveals scattered $s_i^{e-o}$ and $s_i^{o-e}$, which is a typical nature of the multifractal states, as shown in Fig.~\ref{fig:fig2}(b).  
To concretely establish the nature of the state we perform a finite-size scaling analysis of the fractal dimension $D_2$. In Fig.~\ref{fig:fig2}(c), we plot $\ln(\text{IPR}_2)$ versus $\ln(L)$ for some exemplary states from different regions for different system sizes $L$. The slope of each curve yields the associated value of $D_2$. As expected, we obtain $D_2 = 1$ for delocalized states (blue dots), $D_2 = 0$ for the localized states (red diamonds) and $0<D_2<1$ for the multifractal states (green triangles) ~\cite{Fraxnet2022, Roy2022, Sthitadhi2018, Evers2008}. Finally, to visualize the real-space nature of the states we plot $|\psi|^2$ versus the site fraction index $i/L$ in Fig.~\ref{fig:fig2}(d). The upper panel shows a delocalized state with uniform distribution through out the lattice, the middle panel depicts a localized state with exponential localization at a particular site and the lower panel illustrates a multifractal state with a self-similar structure. A more detailed multifractal analysis, including scaling exponents and fractal dimensions, is provided in the Supplementary Material~\cite{supmat}. The above analysis clearly establishes the presence of coexisting mobility edges in the spectrum. 

To complement the above numerical findings, we now provide analytical arguments based on the calculation of Lyapunov exponent $\gamma(E)$ obtained using Avila's global theory~\cite{Avila2015, Prange1982, Simon1985, Marx2017, Sorets1991, Davids1995, Ishii1973, Furstenberg1963, Wang2020, Wang2023, Wang2021, Avila2017, Shuchen2021}.
The Lyapunov exponent is defined as,
\begin{equation}
    \gamma(E) = \lim_{L \to \infty} \frac{1}{L} \int_0^{2\pi} \ln \|T_L(E, \phi)\| \frac{d\phi}{2\pi},
\end{equation}
where $T_L(E, \phi)$ is the cumulative transfer matrix over $L$ sites, and $\|T_L(E, \phi)\|$ denotes its norm. A positive $\gamma(E)$ indicates localized states with a localization length $\xi = 1/\gamma(E)$, while $\gamma(E) = 0$ corresponds to delocalized or critical states. This property of $\gamma(E)$ is used as a tool to discern the presence of any MEs in the system and identify critical energies $E_c$ that demarcate transitions between localized ($\gamma(E) > 0$) and delocalized/critical ($\gamma(E) = 0$) regimes.

The spectral properties of the system are determined by the interplay between the parameters $\lambda$ and $\alpha$, with a critical threshold occurring at $\lambda \alpha = 1$. For $\lambda \alpha < 1$, the quasiperiodic potential remains bounded, facilitating a well-defined localization transition between delocalized and localized states. In contrast, when $\lambda \alpha > 1$, the potential becomes unbounded at certain sites, introducing qualitatively distinct spectral characteristics and fundamentally altering the nature of the eigenstates.
For the bounded regime ($\lambda \alpha < 1$), the critical energies are given by $E_c =  \frac{\pm2 - \lambda}{\lambda \alpha}$.
Localized states ($\gamma(E) > 0$) emerge for energies $|E| > E_c$, with a localization length $\xi = 1/\gamma(E)$, while delocalized states ($\gamma(E) = 0$) prevail for $|E| < E_c$. The Lyapunov exponent in this regime takes the form:
\begin{equation}
    \gamma(E) = \max \left\{ \ln \left( \frac{2f(E)}{1 + \sqrt{1 - (\lambda \alpha)^2}} \right), 0 \right\},
\end{equation}
where $f(E)$ is the dominant eigenvalue magnitude of the transfer matrix, defined as:
\begin{equation}
    f(E) = \frac{|\lambda \alpha E + \lambda| + \sqrt{(\lambda \alpha E + \lambda)^2 - 4(\lambda \alpha)^2}}{4}.
\end{equation}
However, for $\lambda > 0$, the energy $E_c = -\frac{2 - \lambda}{\lambda \alpha}$ lies outside the spectrum when $0 \leq \lambda \alpha < 1$. Consequently, only one mobility edge exists:
\begin{equation}
    E_c = \frac{2 - \lambda}{\lambda \alpha}.
    \label{eq:Eq6}
\end{equation}
In contrast, for the unbounded regime ($\lambda \alpha \geq 1$), the critical energies shift to:
\begin{equation}
    E_c = \pm  2 - \frac{1}{\alpha} .
    \label{eq:Eq7}
\end{equation}

Here, localized states ($\gamma(E) > 0$) occur for $|E| > E_c$, with $\xi = 1/\gamma(E)$, while critical states ($\gamma(E) = 0$) arise for $|E| < E_c$, characterized by multifractal analysis. In this regime, the Lyapunov exponent simplifies to: 
$\gamma(E) = \ln \left( \frac{2f(E)}{\lambda \alpha} \right)$,
which is independent of $\epsilon$. The details for the analysis of the above result can be found in the supplementary material~\cite{supmat}.
To compare the analytically obtained MEs with the numerical ones, we plot the eigen energies as a function of $\lambda$ along with the $D_2$ values of the states for $\alpha=0.2$ in Fig.~\ref{fig:fig1}(b). The lines are obtained from the analytical expressions given in Eq.~\ref{eq:Eq6} and Eq.~\ref{eq:Eq7}. This shows an excellent agreement between the two approaches which confirm the coexistence of an ME (dashed green line) and two AMEs (solid black lines) in a single system.

\begin{figure}[!t]
\centering
\includegraphics[width=1.0\columnwidth]{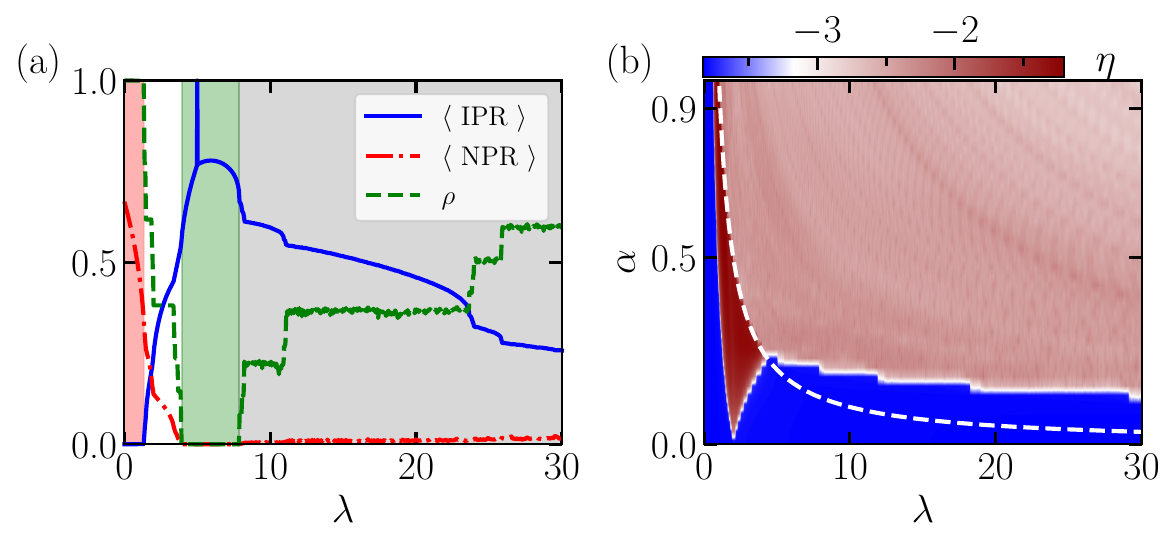}
\caption{(a) $\langle\text{IPR}\rangle$ (blue solid line), $\langle\text{NPR}\rangle$ (red dash-dotted line), and $\rho$ (green dashed line) is plotted as functions of $\lambda$. The white and gray shaded regions correspond to the normal and anomalous intermediate phases, while the red (green) regions indicate fully delocalized (localized) states. (b) Phase diagram in the $\alpha$-$\lambda$ plane showing regions characterized by $\eta$ values: delocalized or localized states (blue region), intermediate phase with conventional ME (dark red region), and with AMEs (light red region). The white dashed line represents the $\lambda \alpha = 1$ boundary, separating bounded and unbounded quasiperiodic potential regimes. All calculations are performed for a system size $L = 6765$ under PBC. }
\label{fig:fig3}
\end{figure}

Now, we aim to separate out the delocalized, localized, and intermediate phases hosting the MEs. To this end we compare $\langle \text{IPR} \rangle$ (blue line) and $\langle \text{NPR} \rangle$ (red line) as a function of $\lambda$ in Fig.~\ref{fig:fig3}(a) for $\alpha=0.2$, where, $\langle \cdot \rangle$ denotes the average over all the states. Up to $\lambda\approx1.325$, the finite $\langle \text{NPR} \rangle$ and vanishing $\langle \text{IPR} \rangle$ denotes the delocalized phase (light pink region). Between $1.325 < \lambda < 3.925$, both $\langle \text{IPR} \rangle$ and $\langle \text{NPR} \rangle$ are finite which indicates the first intermediate phase hosting the conventional ME. After $\lambda > 3.925$, a complete localization of the states occurs indicated by vanishing $\langle \text{NPR} \rangle$ and finite $\langle \text{IPR}\rangle$. However, further increase in $\lambda$ results in an increase in $\langle \text{NPR} \rangle$ although the increase is very small. This increase in $\langle \text{NPR} \rangle$ after $\lambda>7.85$ suggests that some of the localized states turn extended and this is the second intermediate phase which hosts the two AMEs. We highlight that although the value of $\langle \text{NPR} \rangle$ is very small in this region, a large number of localized states become extended. To quantify the fraction  of extended states in the spectrum we define $\rho = \frac{N}{L}$, where $N$ is the number of extended states (with finite $\text{NPR}^{(m)}$) and $L$ is the total number of states. The plot of $\rho$ (green curve) as a function of $\lambda$ in Fig.~\ref{fig:fig3}(a) shows a fully delocalized ($\rho\approx1$) to the first intermediate phase ($1<\rho<0$) to fully localized phase ($\rho\approx0$) and then to the second intermediate phase ($0<\rho<1$). The gradual increase in the value of $\rho$ in the second intermediate phase indicates that the number of localized states becoming extended increases with increase in disorder strength. As already discussed before, the extended states in the second intermediate phase are multifractal in nature. It is important to highlight that the transformation of a large number of localized states into extended states reveals a different kind of re-entrant transition in contrast to the earlier ones where a small fraction of localized states become extended~\cite{Shilpi2021,padhan_prbl_2022}.

To understand the scenarios for all the values of $\alpha$, we obtain a phase diagram by utilising the parameter $\eta$, defined as $\eta = \text{log}_{10} (\langle \text{IPR} \rangle \times \langle \text{NPR} \rangle)$, where, $\langle \cdot \rangle$ denotes the average over all the states. The quantity $\eta$ serves as a quantitative measure to distinguish between localized, delocalized and the intermediate regimes~\cite{padhan_prbl_2022, Shilpi2021}.
In Fig.~\ref{fig:fig3}(b), we plot $\eta$ in $\alpha$-$\lambda$ plane. The blue region before (after) $\lambda=2$ line is the delocalized (localized) phase. The deep red colored region is the intermediate phase where the conventioanl ME appears. The light red region hosts the AMEs and we call this the anomalous intermediate phase.

The emergence of critical states and reentrant behavior can be attributed to the nature of the potential, which is governed by the interplay of two key parameters: $\lambda$ and $\alpha$. The critical threshold $\lambda\alpha = 1$ plays an important role in shaping the system's behavior. For $\lambda\alpha < 1$, the potential remains bounded, leading to a well-defined localization transition between the delocalized and localized states. In contrast, for $\lambda\alpha \ge 1$, the potential becomes unbounded and diverges at specific sites. On the other hand, in this regime, as $\lambda$ increases, the potential becomes weaker. The interplay between this weakening of overall strength and the unbounded nature of the potential gives rise to extended states that exhibit multifractal characteristics.

\begin{figure}[t]
\centering
\includegraphics[width=1.0\columnwidth]{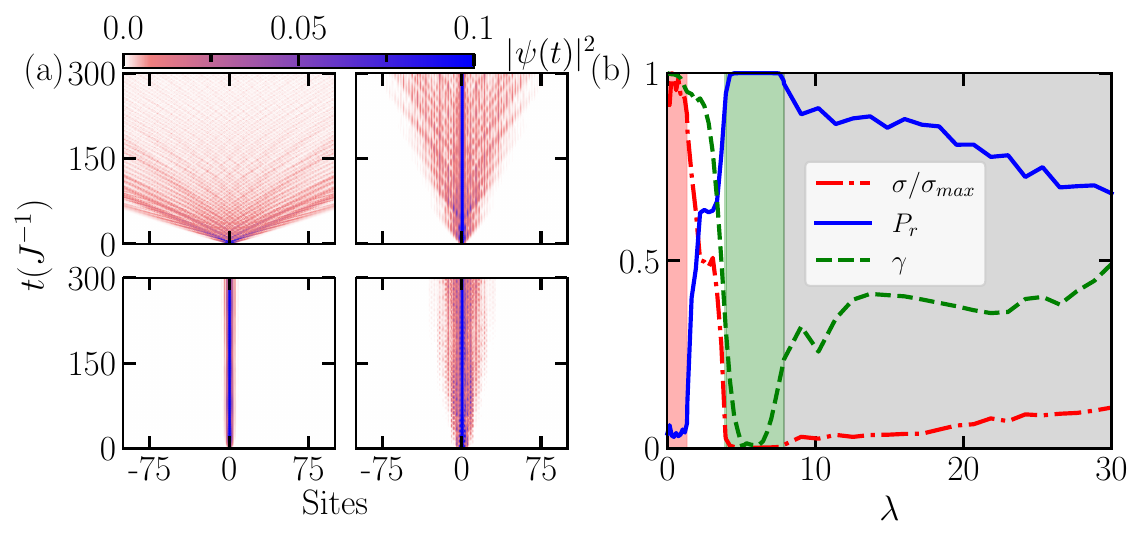}
\caption{(a) Probability density $|\psi(t)|^2$ (color coded) as a function of time $t(J^{-1})$ and site index $i$. The upper left (right) panel shows the result for $\lambda = 0.74$ ($3.07$). The lower left (right) panel shows the result for $\lambda =6.74$ ($25.34$). Here, the data for middle $201$ sites are depicted for clarity. (b) Normalized root-mean-square displacement $\sigma/\sigma_{\text{max}}$ (red dash-dotted line), survival probability $P_r$ (blue solid line), and quantum diffusion exponent $\gamma$ (green dashed line) plotted against $\lambda$ at $t = 10^4 (J^{-1})$. In this case an average over $600$ random phases $\phi$ have been considered. Here, the system size is set to $L=987$ with open boundary conditions (OBC), and the color bar is scaled from $0$ to $0.1$ for better visibility.}
\label{fig:fig4}
\end{figure}

Now we move on to present some of the experimentally accessible quantities from the wavepacket dynamics as the model under consideration can be simulated in existing quantum simulators~\cite{expt_Int_ME_GAA, Immanuel_prl_2018,yoshida2024proposalexperimentalrealizationquantum, Roushan2017, Li2023,Ozawa2019, Meier2016, Gadway2015}. The dynamical properties of the system are explored through the time evolution of an initial state $|\psi(0)\rangle$, governed by the time-dependent Schr\"{o}dinger equation $|\psi(t)\rangle = e^{-i\mathcal{H}t}|\psi(0)\rangle$. Here, $\mathcal{H}$ is the Hamiltonian given in Eq.~\eqref{eq:Eq1}, and we set $\hbar = 1$. The initial state corresponds to a particle localized at the center of the lattice, i.e., $|\psi(0)\rangle = | \cdots c_0^\dagger \cdots \rangle$. To analyze the dynamics, we compute two key quantities: the root-mean-square displacement (RMSD)~\cite{Santos_prl_2019, Xu_2020, Zhang2012PRL, padhan_prbl_2022, Paredesreview_2019}
$\sigma(t) = \sqrt{\sum_{i=0}^{L-1} (i - i_0)^2 |\psi_i(t)|^2}$
and the survival probability $P_r(t) = \sum_{i=-\frac{r}{2}}^{\frac{r}{2}} |\psi_i(t)|^2$, which measures the likelihood of the particle remaining within a spatial range from $-\frac{r}{2}$ to $\frac{r}{2}$~\cite{Santos_prl_2019, padhan_prbl_2022}.

In Fig.~\ref{fig:fig4}(a), we visualize the wave packet dynamics, with color indicating $|\psi(t)|^2$ for $\alpha = 0.2$. For $\lambda = 0.74$, corresponding to the delocalized regime, the wave packet spreads ballistically. In contrast, for $\lambda = 3.07$, representing the intermediate region with a conventional ME, the spreading slows down and exibits a bimodal distribution~\cite{padhan_prbl_2022}. At $\lambda = 6.74$, in the localized regime, the wave packet remains confined, showing no significant spreading. For $\lambda = 25.34$, in the intermediate region with AMEs, the wave packet exhibits slow spreading and a bimodal distribution, indicative of subdiffusive behavior.
In Fig.~\ref{fig:fig4}(b), we plot $\sigma/\sigma_{\text{max}}$ and $P_r$ for $r = 40$, at $t = 10000 (J^{-1})$. We also analyze the time dependence of the RMSD by fitting it to a power law ansatz $\sigma \propto t^\gamma$~\cite{Paredesreview_2019}, where $\gamma$ is the diffusion exponent. This analysis reveals distinct dynamical regimes. In the delocalized regime ($\lambda < 1.325$), $\sigma/\sigma_{\text{max}} \approx 1$, $P_r \approx 0$, and $\gamma \approx 1$, indicating ballistic motion. In the intermediate region with conventional ME ($1.325 < \lambda < 3.85$), $\sigma/\sigma_{\text{max}}$ drops between $0$ and $1$, $P_r$ rises slightly, and $\gamma > 0.5$, signifying superdiffusive behavior. In the localized regime ($\lambda > 3.925$), $\sigma/\sigma_{\text{max}} \approx 0$, $P_r \approx 1$, and $\gamma \approx 0$, reflecting localization. Finally, in the intermediate region with AMEs ($\lambda > 7.85 $), $\sigma/\sigma_{\text{max}}$ slowly increases, $P_r$ decreases, and $0 < \gamma < 0.5$, indicating subdiffusive dynamics. This demonstrates that the system can exhibit all types of dynamical behaviors, namely the ballistic, superdiffusive, localized, and subdiffusive, depending on the disorder strength $\lambda$~\cite{supmat}. The clear distinction between the two intermediate phases underscores the rich and complex nature of the system's behavior, offering novel insights into diverse dynamical phenomena.

\paragraph*{Conclusion.-}
\label{sec:Conclusion}

In this work, we present an exactly solvable quasiperiodic model, which features two distinct types of MEs such as the conventional ME and two anomalous MEs.   Another  important finding in our study is the reentrant behavior at strong disorder strength, i.e., after full localization of the states, a major fraction of the localized states become multifractal as a function of disorder strength. We establish these finding both by numerical exact diagonalization and analytical approaches. We also analyse possible experimental signatures of these phases from the wavepacket dynamics and obtained a rich dynamical behaviour ranging from ballistic to superdiffusive to localized and subdiffusive natures. 

Our findings establish an exact theoretical framework to study the coexistence of multiple mobility edges and reentrant transitions in quasiperiodic systems. Given the recent progress in the experimental front, the model under consideration and the results can be simulated using setups like ultracold atoms in optical lattices~\cite{expt_Int_ME_GAA}. This also opens up directions to uncover other systems exhibiting coexisting MEs including the many-body systems.

\paragraph{Acknowledgement.-} T.M. acknowledges support from Science and Engineering Research Board (SERB), Govt. of India, through project No. MTR/2022/000382 and STR/2022/000023.

\bibliography{ref}

\appendix
\clearpage
\onecolumngrid

\begin{center}
\textbf{\large Supplemental Material for \enquote{Emergence of distinct exact mobility edges in a quasiperiodic chain}}
\end{center}

The supplemental material provides a detailed discussion of the theoretical and numerical analyses supporting the coexisting MEs. We first provide the analytical derivation of the MEs using Avila's global theory. Then, to further explore the behavior of the system at the other regime of the parameter $\alpha$, we examine the numerically obtained eigenspectra for $\alpha = 0.5$. Additionally, we provide further numerical characterization of the multifractal states at $\alpha = 0.2$. The wavepacket dynamics is also analysed to extract the diffusion exponents from the root mean square displacement (RMSD) to characterize the system's transport properties. 

\section{A. A\lowercase{nalytical derivation of mobility edges}}

\subsection{A1. T\lowercase{ransfer Matrix Method}}

To calculate the Lyapunov exponent and analytically derive the mobility edges (ME) and localization transitions in our model, we employ Avila’s global theory. In this section, we provide a detail derivation of
Lyapunov exponents. The discrete Schr\"odinger equation governs the system as,
\begin{equation}
    \psi_{j+1} + \psi_{j-1} + V_j \psi_j = E \psi_j,
    \label{eq:schrodinger}
\end{equation}
where $V_j$ represents the potential at site $j$, given by:
\begin{equation}
    V_j = \frac{\lambda \cos(2\pi \beta j + \phi)}{1 - \lambda \alpha \cos(2\pi \beta j + \phi)},
    \label{eq:potential}
\end{equation}
with $\lambda \in \mathbb{R}$, $\alpha \geq 0$ as a parameter, $\phi$ as the phase, and $\beta \in \mathbb{R}$ as an irrational number. For simplicity, we set the hopping amplitude $J = 1$.

The transfer matrix method can be implemented to compute the Lyapunov exponent. From Eq.~\eqref{eq:schrodinger}, the transfer matrices are defined as:

\begin{equation}
    G_j = \begin{pmatrix} E - V_j & -1 \\ 1 & 0 \end{pmatrix}.
    \label{eq:transfer_matrix}
\end{equation}

The product of these matrices over $L$ sites is expressed as:
\begin{equation}
    T_L(E, \phi) = \prod_{j=0}^L G_j.
    \label{eq:product_transfer_matrix}
\end{equation}
The Lyapunov exponent $\gamma(E)$, which quantifies the asymptotic exponential growth rate of wavefunction amplitudes, is defined as:
\begin{equation}
    \gamma(E) = \lim_{L \to \infty} \frac{1}{L} \int_0^{2\pi} \ln \| T_L(E, \phi) \| \, \frac{d\phi}{2\pi},
    \label{eq:lyapunov_exponent}
\end{equation}
where $\|T_L\|$ denotes the norm of the transfer matrix. Since $\det T_L(E) = 1$, the Lyapunov exponent satisfies $\gamma(E) \geq 0$.

\subsection{A2. R\lowercase{eformulation Using Modified Transfer Matrices}}

For convenience, we introduce a modified transfer matrix $\widetilde{G}_j$, defined as:
\begin{equation}
    \widetilde{G}_j = [1 - \lambda \alpha \cos(2\pi \beta j + \phi)] G_j.
    \label{eq:modified_matrix}
\end{equation}
Expanding this expression yields:
\begin{equation}
    \widetilde{G}_j = \begin{pmatrix} 
        E - (\lambda \alpha E + \lambda) \cos(2\pi \beta j + \phi) & -1 + \lambda \alpha \cos(2\pi \beta j + \phi) \\ 
        1 - \lambda \alpha \cos(2\pi \beta j + \phi) & 0 
    \end{pmatrix}.
    \label{eq:expanded_matrix}
\end{equation}
The corresponding transfer matrix for $\widetilde{G}_j$ is:
\begin{equation}
    \widetilde{T}_L(E, \phi) = \prod_{j=0}^L \widetilde{G}_j.
    \label{eq:modified_product}
\end{equation}
By complexifying the phase ($\phi \to \phi + i\epsilon$) and applying Avila’s global theory~\cite{Avila2015, Prange1982, Simon1985, Marx2017, Sorets1991, Davids1995, Ishii1973, Furstenberg1963, Wang2020, Wang2023, Wang2021, Avila2017, Shuchen2021} for analytic $\text{SL}(2, \mathbb{R})$ cocycles, we derive critical energies $E_c$ that demarcate localized and delocalized or critical regimes. The Lyapunov exponent associated with $\widetilde{T}_L(E, \phi + i\epsilon)$ is:
\begin{equation}
    \widetilde{\gamma}(E, \phi + i\epsilon) = \lim_{L \to \infty} \frac{1}{L} \ln \| \widetilde{T}_L(E, \phi + i\epsilon) \|.
    \label{eq:modified_lyapunov}
\end{equation}
This can also be written as:
\begin{equation}
    \widetilde{\gamma}(E, \phi + i\epsilon) = \frac{1}{2\pi} \int \ln \| \widetilde{T}_L(E, \phi + i\epsilon) \| d\phi.
    \label{eq:lyapunov_integral}
\end{equation}
Substituting into Eq.~\eqref{eq:lyapunov_integral}, we obtain:
\begin{equation}
    \gamma(E, \epsilon) = \widetilde{\gamma}(E, \epsilon) - \frac{1}{2\pi} \int \ln[1 - \lambda \alpha \cos(\phi + i\epsilon)] d\phi.
    \label{eq:substitution}
\end{equation}
The integral in Eq.~\eqref{eq:substitution} evaluates to:
\begin{equation}
    \frac{1}{2\pi} \int_0^{2\pi} \ln \left(|1 - \lambda\alpha \cos(\phi + i\epsilon)| \right) d\phi
    =
    \begin{cases}
        \ln \left( \frac{1 + \sqrt{1 - (\lambda\alpha)^2}}{2} \right), & \text{for } 0 \leq \lambda\alpha < 1 \text{ and } |\epsilon| < \ln | \frac{1 + \sqrt{1 - (\lambda\alpha)^2}}{\lambda\alpha} |, \\
        |\epsilon| + \ln \left( \frac{\lambda\alpha}{2} \right), & \text{for } \lambda\alpha \geq 1.
    \end{cases}
    \label{eq:integral_solution}
\end{equation}

\section{A3. C\lowercase{ase Analysis}}
\subsection{Case 1: $0 \leq \lambda\alpha < 1$}
From Eq.~\eqref{eq:integral_solution}, it follows that $\gamma(E, \epsilon)$ and $\widetilde{\gamma}(E, \epsilon)$ share the same slope with respect to $\epsilon$ when $|\epsilon| < \ln \left| \frac{1 + \sqrt{1 - (\lambda\alpha)^2}}{\lambda\alpha} \right|$.
In the large-$\epsilon$ limit, we find:
\begin{equation}
    \widetilde{T}_L(E, \epsilon) = \prod_{j=0}^L e^{-2\pi i \beta j} e^{i|\epsilon|} \begin{pmatrix} -\lambda\alpha E - \lambda & \lambda\alpha \\ -\lambda\alpha & 0 \end{pmatrix} + o(1).
    \label{eq:large_epsilon_limit}
\end{equation}
According to Avila’s general theory, $\widetilde{\gamma}(E, \epsilon)$ is a convex, piecewise linear function of $\epsilon \in (-\infty, \infty)$. Combined with Eq.~\eqref{eq:integral_solution}, the slope of $\widetilde{\gamma}(E, \epsilon)$ with respect to $\epsilon$ is always $1$. Thus, the Lyapunov exponent can be expressed as:
\begin{equation}
    \widetilde{\gamma}(E, \epsilon) = |\epsilon| + \ln f(E),
    \label{eq:lyapunov_large_epsilon}
\end{equation}
where,
\begin{equation}
    f(E) = \left| \frac{|\lambda\alpha E + \lambda| + \sqrt{(\lambda\alpha E + \lambda)^2 - 4(\lambda\alpha)^2}}{4} \right|.
    \label{eq:f_E}
\end{equation}
Considering the convexity of the Lyapunov exponent, the slope of $\gamma(E, \epsilon)$ may be $1$ or $0$ in the region $0 \leq |\epsilon| < \ln \left| \frac{1 + \sqrt{1 - (\lambda\alpha)^2}}{\lambda\alpha} \right|$. Additionally, the slope of $\gamma(E, \epsilon)$ near $\epsilon = 0$ is nonzero if the energy $E$ lies within the spectrum. Thus, when $E$ is in the spectrum, we have:
\begin{equation}
    \widetilde{\gamma}(E, \epsilon) = |\epsilon| + \ln f(E),
    \label{eq:lyapunov_spectrum}
\end{equation}
for any $\epsilon \in (-\infty, \infty)$. Based on Eq.~\eqref{eq:substitution} and the non-negativity of $\gamma(E, \epsilon)$, we obtain:
\begin{equation}
    \gamma(E, 0) = \max \left\{ \ln \frac{2f(E)}{1 + \sqrt{1 - (\lambda\alpha)^2}}, 0 \right\}.
    \label{eq:gamma_zero}
\end{equation}
The mobility edge is determined by $\gamma(E) = 0$, leading to:
\begin{equation}
    \frac{2f(E)}{1 + \sqrt{1 - (\lambda\alpha)^2}} = 1.
    \label{eq:mobility_edge_condition}
\end{equation}
Solving this equation gives:
\begin{equation}
    |\lambda\alpha E + \lambda| = 2.
    \label{eq:mobility_edge_solution}
\end{equation}
For the bounded regime ($\lambda\alpha < 1$), the critical energies separating localized and delocalized states are:
\begin{equation}
    E_c =  \frac{\pm2 - \lambda}{\lambda \alpha}.
    \label{eq:critical_energy_bounded}
\end{equation}

Localized states ($\gamma(E) > 0$) emerge for energies $|E| > E_c$, with a localization length defined as $\xi = 1/\gamma(E)$. In contrast, delocalized states ($\gamma(E) = 0$) prevail for $|E| < E_c$. However, for $\lambda > 0$, the energy $E_c = -\frac{2 - \lambda}{\lambda \alpha}$ lies outside the spectrum when $0 \leq \lambda\alpha < 1$. Consequently, only one mobility edge exists:
\begin{equation}
    E_c = \frac{2 - \lambda}{\lambda \alpha}.
    \label{eq:single_mobility_edge}
\end{equation}
\subsection*{Case 2: $\lambda\alpha \geq 1$}

The derivation of mobility edges for $\lambda\alpha \geq 1$ parallels the case for $\lambda\alpha < 1$ until Eq.~\eqref{eq:substitution}. For $\lambda\alpha \geq 1$, the result of the integral becomes:
\begin{equation}
    \frac{1}{2\pi} \int \ln |1 - \lambda\alpha \cos(\phi + i\epsilon)| d\phi = |\epsilon| + \ln \left( \frac{\lambda\alpha}{2} \right).
    \label{eq:integral_solution_unbounded}
\end{equation}
Thus, the Lyapunov exponent in the large-$\epsilon$ limit is:
\begin{equation}
    \gamma(E, \epsilon) = \ln \left( \frac{2f(E)}{\lambda\alpha} \right),
    \label{eq:lyapunov_unbounded}
\end{equation}
for any $\epsilon$. Here, $\gamma(E, \epsilon)$ is independent of $\epsilon$. An anomalous mobility edge, separating localized and critical states, is determined by $\gamma(E) = 0$. Solving $(\lambda\alpha E + \lambda)^2 = 4(\lambda\alpha)^2$, we arrive at the exact analytical formula for the anomalous mobility edge:
\begin{equation}
    E_c = \pm 2 - \frac{1}{\alpha}.
    \label{eq:anomalous_mobility_edge}
\end{equation}
In regions where $E > 2 - \frac{1}{\alpha}$ or $E < -2 - \frac{1}{\alpha}$, $\gamma(E) > 0$, and the eigenstates are localized with localization length $\xi = 1/\gamma(E)$. In the region $-2 - \frac{1}{\alpha} < E < 2 - \frac{1}{\alpha}$, multifractal states emerge. 

\vspace{1cm}

\twocolumngrid

\section{B. A\lowercase{nalysis of eigenspectra for $\alpha=0.5$}}
\begin{figure}[t]
\centering
\includegraphics[width=1.0\columnwidth]{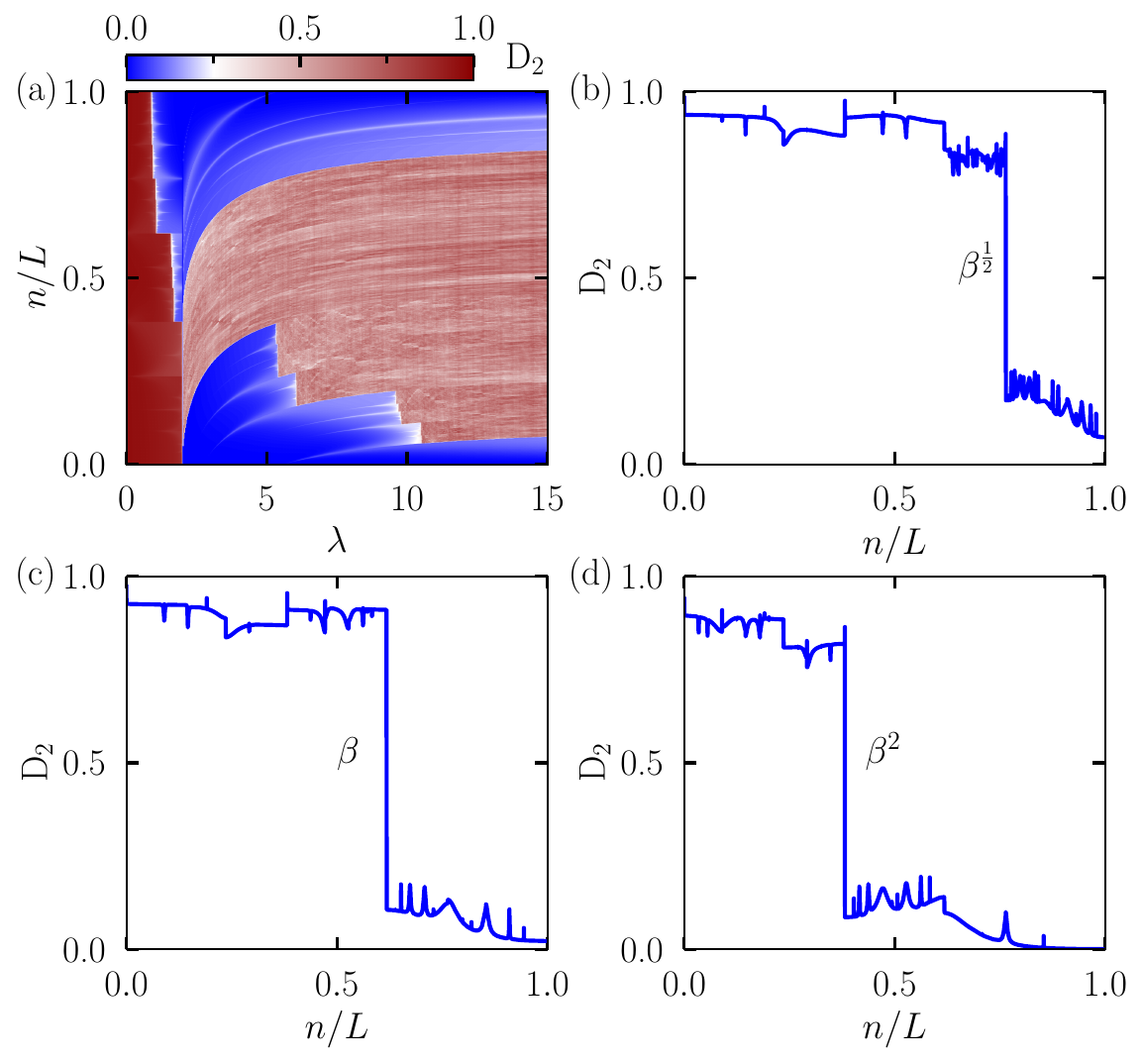}
\caption{(a) Eigenstate index fraction ($n/L$) vs. $\lambda$ with corresponding $D_2$ values for $\alpha = 0.5$. (b-d) $D_2$ as a function of $n/L$ for $\lambda = 1.0, 1.3, 1.85$, showing the relation between the fractions of delocalized states and $\beta^{1/2}$, $\beta$, and $\beta^2$. All results are obtained for a system size $L = 6765$ under periodic boundary conditions (PBC).}
\label{fig:fig5}
\end{figure}

 To further explore the behavior of the system at the other regime of the parameter $\alpha$, in Fig.~\ref{fig:fig5}(a), we present the eigenstate index ($n/L$) as a function of $\lambda$ for $\alpha = 0.5$, alongside the corresponding $D_2$ values. As $\lambda$ increases, the system transitions from a completely delocalized phase to an intermediate regime. Upon further increasing $\lambda$, multifractal states emerge within the intermediate region, separated from localized states by well-defined mobility edges. A notable distinction arises when comparing $\alpha = 0.5$ to $\alpha = 0.2$. For $\alpha = 0.5$, the transitions between the two distinct intermediate phases are significantly sharper, highlighting the potential for novel dynamical behaviors in this regime. Furthermore, at a comparably weaker disorder strength, we observe the presence of the intermediate region with AMEs, where most of the spectrum becomes extended, representing an extreme case of reentrance.

For $\lambda \alpha < 1$, in the intermediate regime, a conventional ME separates the delocalized and localized states. In particular, as disorder strength $\lambda$ increases, the fraction of delocalized states exhibits a distinct stepping structure which is intrinsically linked to the inverse golden mean ratio, $\beta = (\sqrt{5} - 1)/2$, a defining characteristic of quasiperiodicity. This relationship is supported by numerical analysis, with the fraction of delocalized states adhering to the proportions $\beta^{1/2}$, $\beta$, and $\beta^2$ for $\lambda = 1.0$, $1.3$, and $1.85$, respectively, as illustrated in Figs.~\ref{fig:fig5}(b)--(d).

\section{C. Reentrance of multifractal states}

\begin{figure}[t]
\centering
\includegraphics[width=1.0\columnwidth]{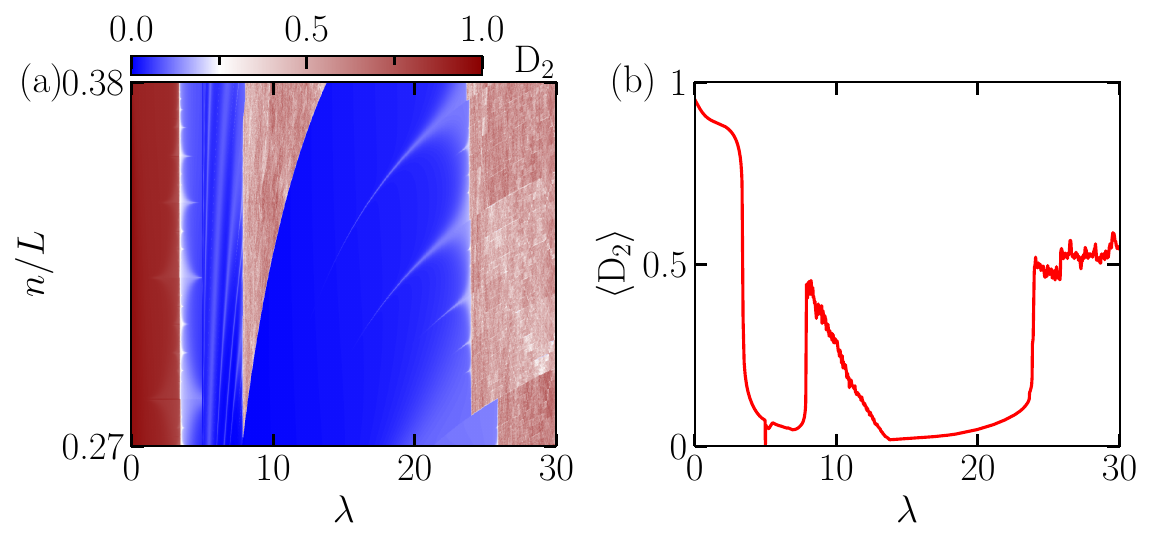}
\caption{(a) Eigenstate index fraction ($n/L$) vs. quasiperiodic strength $\lambda$ in the ratio between $0.27$ to $0.38$, where the color bar shows the corresponding $D_2$ values for $\alpha = 0.2$. (b) An average of $D_2$ over $700$ eigenstates as a function of the $\lambda$ shows re-entrant of the multifractal states.}
\label{fig:fig6}
\end{figure}

Fig.~\ref{fig:fig6}(a) illustrates the eigenstate index ($n/L$) as a function of the quasiperiodic strength $\lambda$ within the range $0.27 \leq \lambda \leq 0.38$, revealing a remarkable re-entrant behavior of multifractal states. Specifically, a subset of the spectrum exhibits a complex sequence of transitions: initially delocalized states become localized, then transition to multifractal states, revert to localized, and finally return to a multifractal regime as $\lambda$ is varied. To substantiate these findings, Fig.~\ref{fig:fig6}(b) displays the average value of $D_2$, calculated over $700$ eigenstates, as a function of $\lambda$. These results provide compelling evidence for the re-entrant nature of multifractal regions, highlighting the intricate spectral transitions driven by the tuning of disorder strength $\lambda$.

\section{D. M\lowercase{ultifractal analysis}}
\label{sec:S_F}

\begin{figure*}[t]
\centering
\includegraphics[width=1.85\columnwidth]{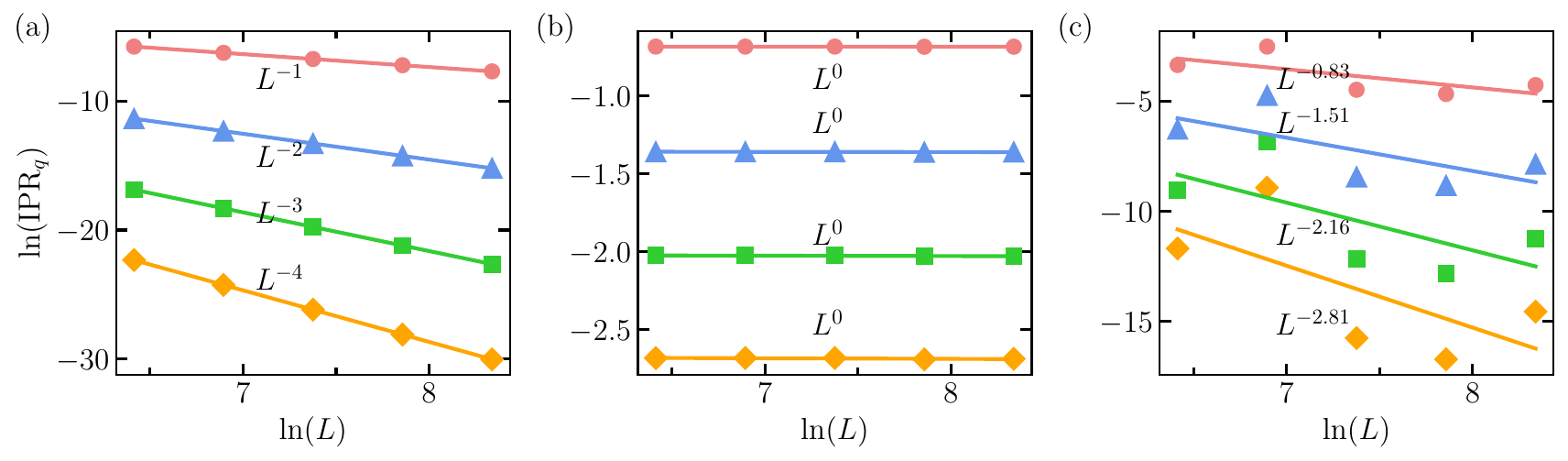}
\caption{The generalized IPR is plotted as a function of system sizes ($L=610, 987, 1597, 2584, 4181$) for different moments $q = 2, 3, 4, 5$,  represented by red circles, blue triangles, green squares, and orange diamonds, respectively. The slopes of the curves, characterized by the mass exponent $\tau_q$, distinguish between (a) delocalized, (b) localized, and (c) multifractal states. Here, the eigenstate index fraction is fixed at $n/L=0.58$ with $\alpha=0.2$.}
\label{fig:fig7}
\end{figure*}

To investigate the multifractal characteristics of eigenstates, particularly those situated between two AMEs, we conduct a comprehensive multifractal analysis. This methodology allows us to elucidate the critical states that emerge within the intermediate regime and distinguish them from delocalized and localized states.
In Fig.~\ref{fig:fig7}, we present the scaling behavior of the generalized inverse participation ratio ($\text{IPR}_q$) for various system sizes $L$.

Fig.~\ref{fig:fig7}(a) depicts $\ln(\text{IPR}_q)$ as a function of $\ln(L)$ for different moments $q = 2, 3, 4, 5$, represented by red circles, blue triangles, green squares, and orange diamonds, respectively. For delocalized states, the data in Fig.~\ref{fig:fig7}(a), exhibit a clear linear dependence on $\ln(L)$, indicative of a uniform spatial distribution throughout the lattice, thus confirming the delocalized nature of these states.
In Fig.~\ref{fig:fig7}(b), the analysis is extended to localized states. The invariance of $\ln(\text{IPR}_q)$ with increasing $L$ reflects the defining feature of exponentially localized wavefunctions, confirming their spatial confinement to a small lattice region irrespective of system size. Finally, in Fig.~\ref{fig:fig7} (c), we analyze the multifractal states. The $\ln(\text{IPR}_q)$ exhibits a nonlinear dependence on $\ln(L)$, reflecting the intricate self-similar structure inherent to these states. The slopes of the fitted lines vary systematically with $q$, providing direct evidence of their multifractal nature. The extracted exponents, such as $L^{-0.83}$, $L^{-1.51}$, $L^{-2.16}$, and $L^{-2.81}$, quantify the fractal dimensions $D_q$ for different moments $q$, offering deeper insights into the spatial complexity of these eigenstates.

\section{E. D\lowercase{ynamical Analysis via} RMSD}
\begin{figure}[t]
\centering
\includegraphics[width=0.8\columnwidth]{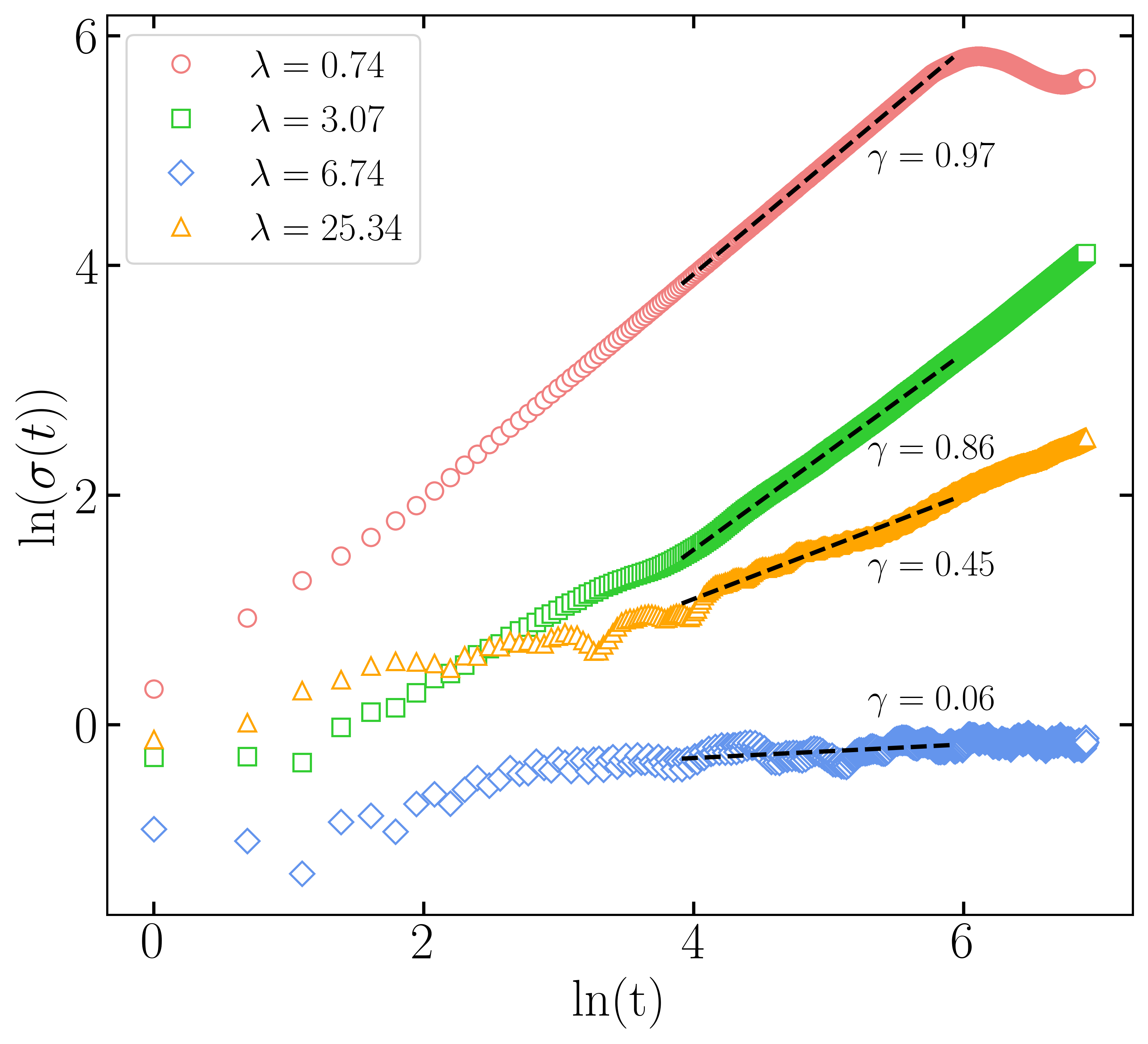}
\caption{Root mean square displacement(RMSD) $(\sigma(t))$ averaged over 600 random phase samples, as a function of time for different regimes: delocalized (red circles), the first intermediate region with ME (orange triangles), localized (blue diamonds), and the second intermediate region with AMEs (yellow squares), corresponding to $\lambda = 0.74, 3.07, 6.74,$ and $25.34$, respectively. The system size is $L = 987$ under open boundary conditions (OBC). The figure also shows the fitting of $\ln(\sigma(t))$ vs. $\ln(t)$, where the slope represents the diffusion exponent $\gamma$ in the time dependence of the RMSD, $\sigma(t) \approx t^{\gamma}$.}
\label{fig:fig8}
\end{figure}

Fig.~\ref{fig:fig8} shows the root mean square displacement $(\sigma(t))$ as a function of time for different dynamical regimes, averaged over 600 random phase samples. This fit is carried out at intermediate time scales, where the transient behavior at short times and the maximum spreading at later times are neglected. The distinct dynamical regimes—delocalized, intermediate with ME, intermediate with AMEs, and localized—are characterized by their respective diffusion exponents $\gamma$, obtained from fitting $\ln(\sigma(t))$ vs. $\ln(t)$. These exponents allow us to identify the ballistic regime ($\gamma \approx 1$, delocalized), the superdiffusive regime ($1/2 < \gamma < 1$, intermediate with ME), the subdiffusive regime ($0 < \gamma < 1/2$, intermediate with AMEs), and the localized regime ($\gamma \approx 0$), associated with the diffusion of an initially localized wave packet. These exponents not only differentiate various transport behaviors across parameters but also uncover distinctive features of two intermediate regions associated with conventional ME and AMEs.

\end{document}